\documentclass{llncs}

\usepackage{graphicx}
\begin{document}

\title{PRIMEBALL: \\a Parallel Processing Framework Benchmark \\for Big Data Applications in the Cloud}

\author{Jaume Ferrarons, Mulu Adhana, Carlos Colmenares, Sandra Pietrowska, \\Fadila Bentayeb and J\'er\^ome Darmont}

\institute{Universit\'e de Lyon (Laboratoire ERIC)\\
Universit\'e Lumi\`ere Lyon 2 -- 5 avenue Pierre Mend\`es-France\\
69676 Bron Cedex -- France\\
first\_name.last\_name@univ-lyon2.fr}
\maketitle

\begin{abstract}

In this position paper, we draw the specifications for a novel benchmark for comparing parallel processing frameworks in the context of big data applications hosted in the cloud. We aim at filling several gaps in already existing cloud data processing benchmarks, which lack a real-life context for their processes, thus losing relevance when trying to assess performance for real applications. Hence, we propose a fictitious news site hosted in the cloud that is to be managed by the framework under analysis, together with several objective use case scenarios and measures for evaluating system performance. The main strengths of our benchmark definition are parallelization capabilities supporting cloud features and big data properties.

\end{abstract}

\textbf{Keywords:} Benchmark, Cloud Computing, Parallel Processing Framework, Big Data, Real Data

\section{Introduction}
\label{sec:Introduction}

%\subsection{Preamble}

We are currently living through an information revolution that has undoubtedly brought a massive increase in the volume of data being produced and stored worldwide. In this Internet age, where the world creates 2.5 exabytes of data every day \cite{bigdata}, traditional approaches
%like TPC-DS \cite{TPCDS} and traditional 
and techniques for data analysis proved limited because some lack parallelism, and most lack fault tolerance capabilities. Therefore, in recent years, many platforms for parallel processing have been created so as to satisfy this need. These platforms provide frameworks for storing, accessing, updating and deleting data efficiently in computer clusters, ensuring fault tolerance and making the whole process transparent to users. Examples of such systems include Google's BigQuery \cite{bigquery} and Apache's Hadoop \cite{hadoop}.

In this context, the terms ``big data'' are used for referring to digital information that comes in high volume, velocity and variety \cite{bigdata}; and the systems that make use of this type of data for achieving profitable objectives can be referred to as big data applications. Several examples of big data applications can be found in the areas of capital market, risk management, retail, social media analysis and meteorology. This kind of applications, beside requiring high parallel processing capabilities for analysis, also needs a good and scalable infrastructure capable of adapting quickly to an increment in computing or storage needs. Therefore, many big data applications are being deployed in the cloud so as to allow fast adaptability and flexibility.

Given the recent increase of big data applications in the cloud, and the use of parallel processing frameworks for dealing with the technical issues implied by the use of clusters and large amount of complex data, it has become important to fix standards so as to allow accurate comparisons of these frameworks. Several benchmarks already exist for measuring a system's parallelization capabilities, cloud features or big data analysis abilities, but none of them offers direct means of accurately measuring: 1) the three of them 2) in a real-life context.

Thus, following the principles defined by Folkerts et al. \cite{cloudbench}, we propose the specifications of PRIMEBALL to position it as a complete and unified benchmark for assessing a system's performance w.r.t. two main axes involved in the context of big data applications hosted in the cloud: parallel processing frameworks and cloud computing service providers. PRIMEBALL also aims to emulate common usages of cloud services, data manipulations and data transfers.

The remainder of this position paper is organized as follows. Section~\ref{sec:relatedwork} reviews existing benchmarks similar to PRIMEBALL and motivates its design. Section~\ref{sec:PRIMEBALLPrinciple} provides an overview of PRIMEBALL. Then, Sections~\ref{sec:PRIMEBALL DATASET}, \ref{sec:PRIMEBALL WORKLOAD} and \ref{sec:PRIMEBALL PROPERTIES AND METRICS} detail the specification of PRIMEBALL's components, i.e., its dataset, workload, and properties and metrics, respectively. Finally, Section~\ref{sec:conclusions} concludes this paper and provides future research leads.

%are a description of the BigData application that serves as a context for PRIMEBALL, section  presents the set of proposed measures for comparing parallel processing frameworks, and finally the last section contains the conclusions and future work.

\section{Related Work}
\label{sec:relatedwork}

%In this section some existing Big Data and Cloud benchmarks are introduced. Their objectives are compared to PRIMEBALL's objectives to explain the novelty of the system presented. First, the different benchmarks are described briefly:

Among the standard TPC benchmarks, TPC-DS, a decision support benchmark that models several generally applicable aspects of a decision support system, including queries and data maintenance \cite{TPCDS}, is closely related to data analytics we target at. However, although it can generate high volumes of data, its underlying business model is a classical retail product supplier, thence its dataset could not fully qualify as big data-oriented because of a lack in structural variety.

MalStone is benchmark for data intensive computing and analysis \cite{MalSton}. It features MalGen, a synthetic data generator that produces large datasets to perform benchmarking. Data is designed to assess systems from the parallel processing point of view. Data is generated probabilistically following specified distributions.
 
Cloud Harmony measures the performance of cloud providers as black boxes \cite{cloudharmony}. The tests performed are mainly focused on assessing hardware performance or specific technologies. Cloud Harmony actually aggregates the results of benchmarks that existed before.

The Yahoo! Cloud Serving Benchmark (YCSB) is a framework to facilitate performance comparisons among cloud database systems \cite{YCSB} that mainly focuses on key-value stores such as Dynamo \cite{Dynamo}. YCSB defines several metrics and workloads to measure the behavior of the systems in different situations, or the same system when using different configurations. 
  
Finally, the Statistical Workload Injector for MapReduce, or SWIM benchmark, is an open source benchmark that enables rigorous performance measurement of MapReduce systems \cite{swim}. It contains suites of workloads of thousands of jobs, with complex data, arrival, and computation patterns, and therefore provides workload-specific optimizations. SWIM is currently integrated with Hadoop.

We provide in Table~\ref{table:benchmarkcomparison} a synthetic comparison of all the above-mentioned benchmarks' properties, as well as PRIMEBALL's as a point of reference.
\begin{center}
    \begin{table}
    \begin{tabular}{ | c | c | c | c | c | c | c |}
    \hline
    &TPC-DS&MalStone&Cloud Harmony&YCSB&SWIM&PRIMEBALL \\ \hline
    Real data&$\sim$&No&$\sim$&No&No&Yes \\ \hline
    Real workload&Yes&No&$\sim$&No&Yes&Yes \\ \hline
    Parallel processing&Yes&Yes&Yes&Yes&Yes&Yes \\ \hline
    Hardware-oriented&No&No&Yes&No&No&No \\ \hline
    MapReduce-oriented&No&No&No&No&Yes&$\sim$ \\ \hline
    Cloud properties&$\sim$&Yes&No&Yes&Yes&Yes \\ \hline
    Complex data&No&No&Yes&No&Yes&Yes \\ \hline
    Big data&$\sim$&Yes&Yes&Yes&Yes&Yes \\ \hline
    Technology-indep&Yes&Yes&Yes&No&No&Yes \\ \hline
    \end{tabular}
    \caption{Comparison of Benchmark Features}
    \label{table:benchmarkcomparison}
    \end{table}
\end{center}

The first property we compare is whether the data processed by a benchmark is produced artificially or extracted from a real environment. Only PRIMEBALL offers the possibility to work with wholly real data, with the aim to better simulate real applications including facets of the problem difficult to emulate when using random distributions to produce data. Although some of the benchmarks used in Cloud Harmony do use real data, most of them are actually processing artificial data. There are also in between positions such TPC-DS which produces data artificially but trying to follow the structure of a real environment.
Proposing a real workload is the second property we selected. A benchmark system bearing this property is executing real-world operations to better simulate a production environment. PRIMEBALL, SWIM and TPC-DS execute only tasks that are closely related to real-world workloads. Moreover, some benchmarks in Cloud Harmony also execute tasks that are common in real environments but not all, for that reason has been marked with a tilde. 

The next property is satisfied by all analyzed benchmarks, i.e., they are all aimed at assessing parallel processing. %All of them are related in some way to parallel processing.
By contrast, Cloud Harmony is the only benchmark that assesses the performance of specific pieces of hardware. For example, it has benchmarks for measuring CPU performance, memory I/O and disk I/O. The other benchmarks can give a notion of the performance of specific parts of the hardware, but are not that specific.

The MapReduce-related property refers to benchmarks aiming at measuring the quality of a system in terms of the performance obtained when executing MapReduce tasks. SWIM is the only benchmark that is uniquely dedicated to MapReduce. However, if PRIMEBALL is implemented using MapReduce tasks, it can measure performance through them too. Cloud properties refer to the prominent features of cloud computing. All benchmarks but TPC-DS and Cloud Harmony are designed to measure properties such as vertical and horizontal scalability, consistency, etc. (cf. Section~\ref{subsec:PROPERTIES AND PERFORMANCE METRICS}). Even though TPC-DS can be used, e.g., to measure the scale up of a distributed SQL database but it is not its purpose. On the other hand, PRIMEBALL has been designed as well to be able to assess this kind of properties. 

Complex data properties describe the benchmarks that are oriented to execute procedures using complex data structures to assess the system under test (SUT). TPC-DS implements a classical data warehouse with numerical and textual values. MalStone only aims to generate a big dataset as a log file and measures system response while processing it, thus the data processed is not complex. The same is true for YCSB, which assesses the performance of key-value stores. Values can be complex, but they are not processed, only stored. The other benchmarks, including PRIMEBALL, include complex data in some of their procedures. Big data properties describe the benchmark systems that involve analytical aspects over large amounts of data. All benchmarks have analytical situations involving large amounts of data. PRIMEBALL has specifically been designed to satisfy this property. TPC-DS has been marked with a tilde because it can be used for this purpose but it depends mainly on the size of data used.

Finally, technology independence describes the systems the are designed to work with several kinds of technologies. YCSB and SWIM do not fulfill this property, because YCSB is oriented to analyze the performance of key-value stores only; and SWIM only assesses MapReduce procedures, and thus only makes sense when the SUT is able to execute them. The other benchmarks, including PRIMEBALL, can be used in environments that are not constrained by a given technology.

\section{PRIMEBALL Overview}
\label{sec:PRIMEBALLPrinciple}

\subsection{Application Model}
\label{sec:BusinessModel}

PRIMEBALL's contextual application is set around New Pork Times: a fictitious on-line information service including international news, current affairs, documentaries, science, health and lifestyle sections. It is constantly updated and available 24-hours all around the world. New Pork Times hosts articles and multimedia documents about the latest news, as well as a large archive of past information.

All of these data are stored in a system called New Pork Times' News Hub (NPT-NH), which resides in a cluster hosted by some cloud service provider. This cluster is managed by a framework for parallel data processing and provides a remote storage that allows the user to access the files in the cluster without having to worry about their distribution in nodes. This storage system allows the user to insert and update data, and also to execute batch processes for analyzing/processing the data.

\subsection{PRIMEBALL Features}
\label{sec:PRIMEBALLFeatures}

This section lists what PRIMEBALL is/does and does not, so that its position, notably w.r.t. state of the art benchmarks (Section~\ref{sec:relatedwork}), is clear.

On one hand, PRIMEBALL:
    \begin{itemize}
        \item is a benchmark. It aims to compare the performance of the parallel processing framework under test with respect to several meaningful metrics;
        \item is cloud-oriented. The obtained results could also be used to compare:
        \begin{itemize}
            \item cloud platforms as parallel processing frameworks,
            \item service providers executing systems using the same cloud platform;
        \end{itemize}
        \item is repeatable. All the proposed experiments are designed to lead to the same results if they are executed under the same conditions;
        \item is portable. The benchmark has been designed to be implemented in different cloud platforms.
        \item does define a set of operations that is meaningful in the context of parallel processing and cloud computing;
        \item does define performance metrics that are oriented to measure cloud properties. The criteria to assess each metric are also defined;
        \item does define data relationships. We provide a description of the information stored in the SUT to be processed during the benchmark run.\\
    \end{itemize}
    
On the other hand, PRIMEBALL does not:
\begin{itemize}
    \item define technical execution details. It defines guidelines, but given that SUTs can be very different, the relevance of results is tightly related to implementation details;
    \item define expected performance results. No absolute value is provided as a comparison point, given that they are subject to implementation details;
    \item compare data retrieval or processing algorithms;
    \item define a storage schema. However, we define how data are physically stored.
\end{itemize}

Thence, PRIMEBALL is the first cloud-oriented unified benchmark aiming to assess all the elements involved in cloud-based big data application systems.

\section{PRIMEBALL Dataset}
\label{sec:PRIMEBALL DATASET}

For using PRIMEBALL, it is necessary to implement NPT-NH (Section~\ref{sec:BusinessModel}). Therefore, the following subsections contain a technical description of its architecture, the type of data it contains and the operations performed onto data by means of batch processes, such as metadata extraction.

\subsection{Types of Files to be Hosted}

The system's database shall hold only three types of files. However, there can be many files of the same type. The three types of files follow.
\begin{itemize}
\item{General information (XML): This set of files comprise the many XML documents that describe the standard information stored by NPT-NH, i.e., information about authors, the actual news, and so on. Section \ref{sec:schema} describes the conceptual schema of this information.}
\item{Media files (binary): Some articles make references to these files, which can be either audio or video documents.}
\item{Metadata (XML): Several metadata for information retrieval and other tasks are extracted from the other two types of files by internal algorithms, for further querying. These metadata must be persisted as XML files in the system.}
\end{itemize}

\subsection{PRIMEBALL Schema}
\label{sec:schema}
The system must hold as XML files data about the following entities:
\begin{itemize}
    \item{articles: the actual news articles;}
    \item{topics: the topics an article may belong to;}
    \item{keywords: sets of words that roughly describe the content of an article;}
    \item{languages: marks for indicating what language/dialect an article is written in;}
    \item{authors: people who write the articles;}
    \item{journalists: authors who work in journals and make interviews;}
    \item{professionals: specialists in some topic who write special analyses;}
    \item{countries: information about countries authors might be citizens of or work in;}
    \item{dates: information about the day of the year when an article was written;}
    \item{media: reference to a media file with some internal comments.}
\end{itemize}

\begin{figure}[htb]
    \centering
    \includegraphics[width=\textwidth]{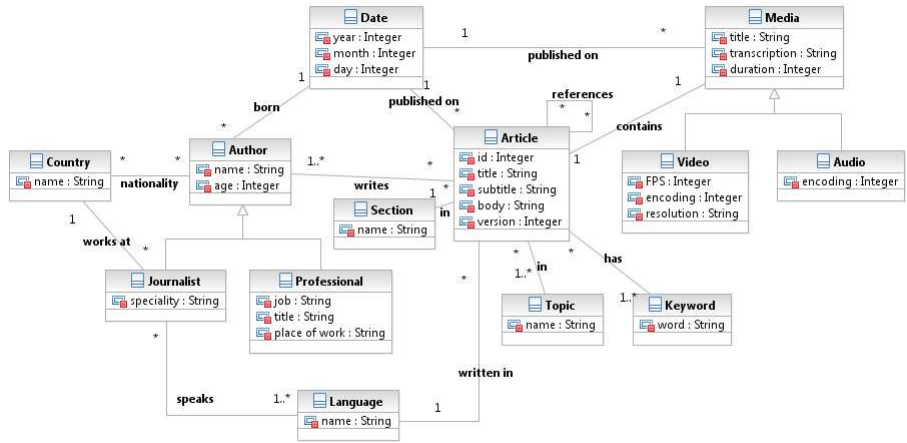}
    \caption{Conceptual Schema of PRIMEBALL's Dataset}
    \label{fig:fig1_conceptial_schema}
\end{figure}

The conceptual schema of this dataset is featured in Figure~\ref{fig:fig1_conceptial_schema}. Its actual implementation depends on the framework for parallel data processing to be benchmarked and its capabilities for storing data.

\subsection{Initial Data}
\label{sec:init}

To create an initial corpus for populating NPT-NH, PRIMEBALL shall come bundled with a crawler that extracts, transforms and loads information from one or more real world news hub akin to New Pork Times (a famous news site may come to mind). The crawler fetches information about news published during a requested period of time, which is recommended to be set up as the last 40 years. Moreover, it must also extract information about authors, media files available, and metadata about the articles relevant to the system's architecture. Due the big data properties of the benchmark, it is recommended to fetch at least 100~TB of data for running the tests. Although, depending on the environment to be benchmarked it might be required to use higher amounts of data. The benchmark can virtually scale up to 1~PB.

Once the corpus has been fetched, it can be sliced at will for selecting any scale factor for the initial data population. However, it is important to bear in mind that if the whole corpus is used in the initialization phase, then it will not be possible to perform updates or scale operations, since there will not be extra data available.

\subsection{Metadata Processes}
\label{sec:metadata}

When data are loaded, it is necessary to run algorithms that extract some metadata from the files and build the structures for the following information retrieval tasks performed by NPT-NH:
\begin{itemize}
\item{run a Hidden Markov Model \cite{HMM4SR} speech recognition algorithm on media files for transcribing the speech to text;}
\item{compute the TF-IDF measure for all articles and transcriptions, as specified in the CDI IDF algorithm \cite{TFIDF};}
\item{compute the page rank of articles, as specified in the weighted pagerank algorithm \cite{pagerank};}
\item{perform a topic extraction routine on articles and transcriptions, as specified in the Latent Dirichlet Allocation algorithm \cite{topicmodels}.}
\end{itemize}

These algorithms have to be implemented as described in the cited references and adapted to be run in the cluster managed by the parallel data processing framework. 

\subsection{Data Scaling and Maintenance}
\label{sec:scaling}

This section describes the processes performed for updating and scaling data in NPT-NH. Basically, these processes are executed integrally as batch tasks. New Pork Times' authors write new articles every day. Articles have to be placed in NPT-NH for allowing access to users. New authors may also come in. The following scaling tasks have to be performed by the system. New data has to be obtained as progressive slices of PRIMEBALL's corpus, making sure that the data was not already inserted. Once new data are inserted, the system has to extract the necessary metadata from them so as to ensure that subsequent queries can be performed.

More specifically, when new information is inserted, it is necessary to recalculate some structures used for information retrieval. Thus, PRIMEBALL contains an implemented procedure for recomputing both TF-IDF metrics from all the documents and topics for all articles.

\section{PRIMEBALL Workload}
\label{sec:PRIMEBALL WORKLOAD}

\subsection{Query Set}
\label{subsec:queries}

This section contains the set of queries that are typically performed by NPT-NH users and that must be used for testing the performance of any given parallel data processing framework.
These queries must be performed over the dataset defined in Section~\ref{sec:PRIMEBALL DATASET} and involve the most important aspects of performance. Although the following set of queries is not finite, it covers a wide range of classes from Figure~\ref{fig:fig1_conceptial_schema}, the relationships between them; it is applicable and enables to successfully measure performance.

One first subset of queries concerns the most common and hottest topics published in the system. The user might be interested in this information in order to know what kind of issues are the most frequently described in a certain interval of time, and therefore attract attention.
\begin{enumerate}
    \item Articles containing the most frequent bigrams, sorted by pagerank.
    \item Articles published during a time interval $I$, sorted by topics. The output contains pairs of article title and topic.
    \item Most frequently used words by a journalist $J$ from each country in the world during a certain time interval $I$.  The output is compound by a list of countries, each one with keywords.
    \item Rank of the keywords used in the articles published on an exact date $D$.
    \item Most frequently used keywords in month $M$ and year $Y$.
\end{enumerate}

Furthermore, in order to evaluate how topics evolve w.r.t. time, it is necessary to include the various time measures in a second subset of queries.
\begin{enumerate}
\setcounter{enumi}{5}
    \item Most frequently used keywords on day $D$ and two different years $Y_1$ and $Y_2$, sorted by descending count.
    \item Articles published on date $D$ with the greatest number of references to the previous year.
    \item Articles related to a topic $T$ very frequently referenced lately, i.e., in a time interval $I$ ending at the current date.
    \item Journalists and professionals who wrote an article on the same month, on the same topic, sorted by days.
\end{enumerate}

Finally, to analyze the diversity of articles and compare them w.r.t. their source, one should consider the following, eventual group of queries.
\begin{enumerate}
\setcounter{enumi}{9}
    \item Articles written by $X$ journalists in a specific time interval $I$ that have at least $Y$ common topics.
    \item Rank of the languages in which articles are written. The output consists in pairs (language, number of articles).
%    \item Journalists and professionals who wrote an article on the same month on the same topic and sort by day. DOUBLE
    \item Articles written by an author $A$ from a given country $C$ that best match a search term $S$.
    \item Documents that best match another document published on the same day and month, but one year later.
    \item Articles that focus on the same topic, but have been written by different journalists who were born before and after a given year $Y$.
\end{enumerate}

\subsection{Test Protocol}
\label{subsec:TEST PROTOCOL}

We present in this section different scenarios to help actually benchmark a system. To define them, a default scale factor $SF$ is used. A dataset of scale factor $SF$ is a set of articles and corresponding metadata having a total size of $SF$~GB.

\subsubsection{Scenario 1}

This scenario simulates the evolution of the system along time, in terms of data operations and queries.\\

\textit{Initial state}
The system contains a dataset with a specified scale factor $SF$.\\

\textit{Operations}
\begin{enumerate}
\item Execute queries 4, 7 and 14 from the generic query set (Section~\ref{subsec:queries}) choosing as date, e.g., September 12, 2001.
\item Double the volume of the dataset according to scenario 7.
\item Repeat the queries executed in step 1, for the same date and another one, e.g., November 5, 2008.
\end{enumerate}

\subsubsection{Scenario 2}

This scenario simulates an extreme situation: a very famous article has been published with many mistakes and publishers are correcting it constantly. Their main concern is to deliver a consistent view of the article to people. Here, we are interested in measuring how many times the article is read in the older versions after being read once in the new version.\\

\textit{Initial state}
The system contains a dataset with a specified scale factor $SF$.\\

\textit{Operations}
\begin{enumerate}
\item Initiate a thread performing 100 queries per second to retrieve the given article.
\item Start another thread updating the same article every 5 seconds.
\end{enumerate}

\subsubsection{Scenario 3}
This scenario simulates node failures in terms of network reachability. Here, we are interested in knowing how many nodes can be removed from the cluster before some data become unreachable.\\

\textit{Initial state}
The system contains a dataset with a specified scale factor $SF$.\\

\textit{Operations}
\begin{enumerate}
\item Execute all queries from the generic query set (Section~\ref{subsec:queries}) sequentially.
\item Remove a node and reiterate step 1.
\end{enumerate}

\subsubsection{Scenario 4}
This scenario aims to measure the concurrency offered by the system while accessing data.\\

\textit{Initial state}
The system contains a dataset with a specified scale factor $SF$.\\

\textit{Operations} 
Execute the following process 300 times and count the number of inconsistencies.
\begin{enumerate}
    \item Start 10 threads executing the whole generic query set  (Section~\ref{subsec:queries}).
    \item {Start 5 threads each of them performing:
\begin{enumerate}
    \item updates in articles: 10 per second;
    \item removing articles: 10 per second;
    \item adding articles: 10 per second.
\end{enumerate}
}
\end{enumerate}

\subsubsection{Scenario 5}
The objective of this scenario is to simulate analysis procedures over the dataset.\\

\textit{Initial state}
The system contains a dataset with a specified scale factor $SF$.\\

\textit{Operations}
Execute queries uniformly selected from the following.
\begin{enumerate}
    \item Top 10 articles seen each month in 2010.
    \item Average number of pages per article for each journalist in the system.
    \item Average age of publishers and standard deviation.
    \item Maximum number of versions of an article.
\end{enumerate}

\subsubsection{Scenario 6}

The aim of this scenario is to initialize the system and make it ready for handling the information for New Pork Times.\\

\textit{Initial state}
Empty storage system.\\

\textit{Operations}
Execute the steps described in Section \ref{sec:init} to initialize the environment.

\subsubsection{Scenario 7}

This scenario is used to increase the volume of the dataset to simulate the fact that new articles are inserted over time.\\

\textit{Initial state}
NPT-NH has a consistent state.\\

\textit{Operations}
Execute the steps described in Section \ref{sec:scaling} for data scaling and maintenance.

\section{PRIMEBALL Properties and Metrics}
\label{sec:PRIMEBALL PROPERTIES AND METRICS}

\subsection{Properties and Performance Metrics}
\label{subsec:PROPERTIES AND PERFORMANCE METRICS}

This section presents the metrics that we use to evaluate the performance of the SUT. We also define the different system properties that can be assessed using PRIMEBALL.

We first specify two main metrics. The first one is throughput. The throughput of the SUT for a given scenario is the total time required to execute it (scenarios are defined in Section~\ref{subsec:TEST PROTOCOL}). The second metric, price performance, takes price (Section~\ref{subsec:PRICING}) into account and is expressed as follows.
$$\textit{Price performance} = \frac{\textit{Throughput}}{Price}$$

Moreover, the set of operations that can be executed against the system in the context of New Pork Times is defined as follows.
\begin{itemize}
\item Read: Obtain one or more articles.
\item Write: Create a new article or a new version, add new journalists, languages, topics...
\item Update: Modify an existing article within the same version or modify the information related to a journalist.
\item Delete: Remove inappropriate content.
\item Search: Obtain articles by matching a search (a set of given words, topics, authors, dates...).
\end{itemize}

Using all these definitions, we can set the following properties.

\subsubsection{Generic Cloud Properties}
\begin{enumerate}
    \item {\textbf{Scale up}: ability of the system to handle more data when adding more computers while maintaining performance.
\begin{itemize}
        \item \textit{Importance}: In the case of a news website, it is very important to be able to scale up the system. There are a lot of news added every day and the service must keep on performing the same.

        \item \textit{Measurement}: To measure this property, scenarios 4 and 5 (Section~\ref{subsec:TEST PROTOCOL}) must be executed twice, doubling the amount of data ($SF$) and the amount of nodes in the cluster the second time. Throughput increase ratio is the metric recommended for this property. $$\textit{Throughput increase ratio} = \frac{\textit{Throughput after}}{\textit{Throughput before}}$$
\end{itemize}
}

    \item{\textbf{Elastic speedup}: adding more computers to the cluster with the same amount of data results in better performance.
\begin{itemize}
    \item \textit{Importance}: For New Pork Times, it is very relevant to know whether the system can offer a better performance when required, e.g., when there is a worldwide event with more people involved than usual looking for news and information. Thus, it is crucial to be able to maintain the quality of service even during peak demands.

    \item \textit{Measurement}: To measure this property, we propose to execute scenarios 2, 4 and 5 (Section~\ref{subsec:TEST PROTOCOL}) in order to observe throughput with the default cluster $SF$ size. The metric we propose is also throughput increase ratio. $$\textit{Throughput increase ratio} = \frac{\textit{Throughput after}}{\textit{Throughput before}}$$
\end{itemize}
}

    \item{\textbf{Horizontal scalability}: ability of the system to distribute evenly the data load and workload among cluster nodes.
\begin{itemize}
    \item {\textit{Importance}: It is very useful to know up to what point one can exploit the current cluster and keep throughput in between some boundaries. In other terms, we determine what highest price performance can be achieved. It is very interesting in two senses:
\begin{itemize}
    \item upper bound: to answer the question ``how many articles can the news website add into the system while keeping response time below 0.2 seconds";

    \item lower bound: to optimize resource usage while fixing a performance lower bound. It might indeed be possible to reduce the number of nodes and offer the same user experience (response time).
\end{itemize}
}

    \item \textit{Measurement}: To assess this property, scenarios 4 and 5 (Section~\ref{subsec:TEST PROTOCOL}) must be executed and system throughput measured. Then, $SF$ is increased and the process repeated. Again, throughput increase ratio can be used to evaluate this property. The closer it is to 1, the better is horizontal scalability. $$\textit{Throughput increase ratio} = \frac{\textit{Throughput after}}{\textit{Throughput before}}$$
\end{itemize}
}

    \item{\textbf{Latency}: time to execute a set of operations.
\begin{itemize}
    \item \textit{Importance}: For New Pork Times, it is essential to be able to show news very quickly to users. If it takes too much time, users are going to look for a different website, thus a low latency is required.

    \item \textit{Measurement}: Latency of the SUT can be measured as the throughput when executing scenarios 4, 5 and 6  (Section~\ref{subsec:TEST PROTOCOL}).
\end{itemize}
}

\vspace{0.25cm}
    \item{\textbf{Durability}: ability of the system to retain information for a long period of time.
\begin{itemize}
    \item \textit{Importance}: In the case of a news website, it is very important to ensure that no information is lost. Users have to be able to check and find information they have read previously.

    \item \textit{Measurement}: {Scenario 1 (Section~\ref{subsec:TEST PROTOCOL}) is intended to measure data durability. We define the durability ratio as a metric for this purpose.
    $$\textit{Durability ratio}=\frac{\textit{Correct reads}}{\textit{Total reads}}$$
}
\end{itemize}
}

    \item{\textbf{Consistency and version handling}: two different readings of the same data at the same time should return the same value.
\begin{itemize}
    \item \textit{Importance}: It is important for a website to give a consistent view of data to all users at the same time around the world. In the proposed model, There may be several revisions of an article, which has to be consistent for all readers.

    \item {\textit{Measurement}: Using scenario 2 (Section~\ref{subsec:TEST PROTOCOL}), the performance of the system for this property can be measured using the consistency ratio as a metric. $$\textit{Consistency ratio}=\frac{\textit{Consistent reads}}{\textit{Total reads}}$$
}
\end{itemize}
}

    \item{\textbf{Availability}: data is accessible even when there are some inaccessible nodes.
\begin{itemize}
    \item \textit{Importance}: It is very relevant for New Pork Times to guarantee the access to all the news stored in the system.

     \item \textit{Measurement}: Scenarios 3 and 6 (Section~\ref{subsec:TEST PROTOCOL}) aim to measure this property, thus the throughput of the SUT can be taken as a metric for this property.
\end{itemize}
}

\vspace{0.25cm}
\item{\textbf{Concurrency}: the system has to be able to offer a service to different clients at the same time.
\begin{itemize}
    \item \textit{Importance}: In New Pork Times, users can keep reading while publishers are adding news, and the system has to be able to handle the multiple operations of different natures at the same time.

    \item{\textit{Measurement}: Given concurrent scenario 4 (Section~\ref{subsec:TEST PROTOCOL}), we propose two metrics:
\begin{itemize}
    \item system throughput;
    \item concurrency ratio. $$\textit{Concurrency ratio}=\frac{\textit{Successful operations}}{\textit{Total operations}}$$
\end{itemize}
}
\end{itemize}
}
\end{enumerate}

\subsubsection{Complex Data Properties}
\begin{enumerate}
\setcounter{enumi}{8}
\item {\textbf{Path traversals}: ability of the system to link data from different parts of the schema using the defined relationships.
\begin{itemize}
    \item \textit{Importance}: In the case of a news Web site environment, this property is very important to improve search experience.

    \item \textit{Measurement}: Queries 3, 4, 7, 10 and 12 (Section~\ref{subsec:queries}) from the generic query set involve following a path through different class relations to link concepts. The throughput of this type of queries is used to measure that property.
\end{itemize}
}
\item {\textbf{Construction of complex results}: ability of the system to generate (semi)-structured output from the information system.
\begin{itemize}
    \item \textit{Importance}: This property is very relevant to a news website, mainly to allow analysis over the contained data.

    \item \textit{Measurement}: The generic query set defined in Section~\ref{subsec:queries} contains queries with complex results, i.e., queries 2, 3, 6 and 11. The throughput of these queries can be used as a metric for this property. Moreover, scenario 6 (Section~\ref{subsec:TEST PROTOCOL}) has to be used to measure this property.
\end{itemize}
}
\item {\textbf{Polymorphism}: ability of the system to deal with type inheritances, i.e., treating types and subtypes of objects to compute query results.
\begin{itemize}
    \item \textit{Importance}: Inheritance is a good way to deal with complex relationships between objects. For this reason, the performance of the system while executing these kinds of operations is very relevant.

    \item \textit{Measurement}: Fix a cluster and an initial workload, then execute and measure system performance while executing queries 3, 9, 12 and 15 (Section~\ref{subsec:queries}; all of them involve inheritance operations).
\end{itemize}
}
\end{enumerate}

\subsubsection{Big Data Properties}
\begin{enumerate}
\setcounter{enumi}{11}
\item {\textbf{Analysis}: ability of the system to generate summarized data and statistical information.
\begin{itemize}
    \item \textit{Importance}: For a news website, having statistics such as how many times an article has been read, average words per article, etc., is very relevant.
    \item \textit{Measurement}: This property can be measured in terms of throughput while executing analytical scenario number 5 (Section~\ref{subsec:TEST PROTOCOL}).
\end{itemize}
}
\end{enumerate}

\subsubsection{Information Retrieval Properties}
\begin{enumerate}
\setcounter{enumi}{12}
\item {\textbf{Full text}: being able to search a single word in all documents simultaneously.
\begin{itemize}
    \item \textit{Importance}: This property is very relevant to a news website to allow users searching information easily in the system.
    \item \textit{Measurement}: It can be measured in terms of throughput when searching for different terms, some famous, some normal and some strange, e.g., Obama, Higgs, Star Trek, Cleopatra, etc. Queries of this type are included in the generic query set (Section \ref{subsec:queries}).
\end{itemize}
}
\end{enumerate}

\subsection{Pricing}
\label{subsec:PRICING}
In Section \ref{subsec:PROPERTIES AND PERFORMANCE METRICS}, we defined sytem performance w.r.t. time and cost. The main pricing factors involved in processing data in the cloud follow.
\begin{itemize}
    \item Cloud provider: different cloud service providers may have different pricing policies.
    \item Number of instances and type: infrastructure used to execute PRIMEBALL.
    \item Required storage space: it is directly related to the scale factor used (Section~\ref{subsec:TEST PROTOCOL}).
    \item Platform inherent costs: operation, administration and maintenance.
    \item Execution time: time spent to run the tests.
\end{itemize}
The specific cloud provider model has to be applied to compute real cost.

\section{Conclusions}
\label{sec:conclusions}

We propose in this paper the specifications for PRIMEBALL, a complete and unified benchmark for measuring the characteristics of parallel cloud processing frameworks for big data applications. In front of the already existing benchmarking options, PRIMEBALL can be used as a guideline to build an integral solution for benchmarking could platforms. The real-life model adopted in PRIMEBALL is that of a fictitious news hub called New Pork Times, which is basically a fair approximation of a popular real-life news site. The general architecture and inner processes of the system are well-defined so as to allow an unambiguous implementation of the benchmark. 

The workload applied on this news dataset is not only made of queries, but also of data-intensive 
batch processes. Moreover, we propose several use-case scenarios together with relevant metrics for assessing the framework's performance from different points of view, such as data availability or horizontal scalability. The novelty of our work lies in the fact that existing, related benchmarks measure parallelization capabilities, cloud features, big data analysis ability, but none of them combines all these properties while exploiting real-life data. 

Future work on PRIMEBALL will be mainly focused on implementing a crawler for fetching and transforming real data from the Web to feed the benchmark's dataset. Distributing the built dataset online will improve the repeatability of the experiments.  Moreover, the actual feasibility and relevance of PRIMEBALL shall be validated by actually implementing the benchmark in several cloud environments to obtain experimental results and by publishing performance comparison results. For instance, implementation in popular data processing frameworks such as Hadoop should be achieved.

Moreover, future extensions of the benchmark could include new scenarios that exploit different properties of cloud providers, such as vertical growth of the cluster, or new measures such as efficiency of bandwidth use. Actual experiments should also help refine the benchmark's workload.

\pagebreak

\end{document}